\documentclass[preprint,12pt]{elsarticle}

\usepackage{amsmath}
\usepackage{amsfonts}
\usepackage{url}
\usepackage{color}
\usepackage{xcolor}
\usepackage{graphicx}
\usepackage{stmaryrd} 
\usepackage{cancel}
\usepackage{amssymb}

\DeclareMathOperator{\Tr}{Tr}
\definecolor{darkcyan}{rgb}{0.0, 0.55, 0.55}

\newcommand{\PGR}[1]{{\color{black} #1}}

\newcommand{\MD}[1]{{\color{black} #1}}

\journal{Annals of physics}

\begin{document}

\begin{frontmatter}



\title{Stochastic TDHF in an exactly solvable model}


\author{L.~Lacombe}
\address{Laboratoire de Physique Th\'eorique, Universit\'e de Toulouse, CNRS, UPS, France}
\author{E.~Suraud}
\address{Laboratoire de Physique Th\'eorique, Universit\'e de Toulouse, CNRS, UPS, France}
\author{P.-G.~Reinhard}
\address{Institut f{\"{u}}r Theoretische Physik, Universit{\"{a}}t Erlangen,
             Staudtstra\ss e 7, D-91058 Erlangen, Germany}
\author{P. M. ~Dinh\footnote{corresponding author :  dinh@irsamc.ups-tlse.fr}}
\address{Laboratoire de Physique Th\'eorique, Universit\'e de Toulouse, CNRS, UPS, France}

\begin{abstract}
{We apply in a schematic model a theory beyond mean-field, namely
Stochastic Time-Dependent Hartree-Fock (STDHF), which includes dynamical elec\-tron-elec\-tron
collisions on top of an incoherent ensemble of mean-field states by occasional 2-particle-2-hole ($2p2h$) jumps.
The model considered here is inspired by a Lipkin-Meshkov-Glick model of $\Omega$ particles 
distributed into two bands of energy and coupled by a two-body interaction. 
Such a model can be exactly solved (numerically though) for small $\Omega$. It therefore allows a direct
comparison of STDHF and the exact propagation. The systematic impact of the model parameters as
the density of states, the excitation energy and the bandwidth is presented and discussed.
The time evolution of the STDHF compares fairly well with the exact entropy, as soon as the excitation
energy is sufficiently large to allow $2p2h$ transitions.
Limitations concerning low energy excitations and memory effects are also discussed. 
}
\end{abstract}


\end{frontmatter}



\section{Introduction}

Time-dependent mean-field methods are widely used tools to describe
the dynamics of many-fermion systems, for example in the framework of
time-dependent density functional theory in electronic systems
\PGR{\cite{Gro96,Mar04,Fen10}} or of time-dependent Hartree-Fock (TDHF) in nuclei
\PGR{\cite{Dav85a,Sim12,Mar14}}. However, at high excitations and/or over long
simulation times, dynamical correlations, neglected in mean-field
propagation, become increasingly important. These have been studied
extensively in homogeneous systems as quantum liquids
\cite{Kad62,Bal75}. Dynamical correlations for finite systems are much
more demanding and have been treated mostly in semi-classical
approximation by the Vlasov-Uehling-Uhlenbeck (VUU) approach which
has found wide spread application, e.g., in nuclear physics
\cite{Ber88,Dur00}, in laser excitation of metal clusters
\cite{Dom98b,Fen04,Fen10}, or in electron transport in wires
\cite{Nit03a}. A fully quantum-mechanical description of dynamical
correlations in finite fermion systems is much more demanding. One
promising line of development is Stochastic TDHF (STDHF) where
correlations are handled in terms of an ensemble of mean-field states
generated by stochastic jumps into 2-particle-2-hole 
states~\MD{\cite{reinhard_stochastic_1992,Lac14}}. Recently, first practical tests came
up in one-dimensional many-\MD{electron} systems \cite{Sur14a,Sla15a}. 

The aim of this paper is to continue testing of STDHF by comparison
with an exact solution. To this end, we employ a sufficiently simple
schematic model.  Starting point is the Lipkin-Meshkov-Glick (LMG)
model
\cite{lipkin_validity_1965,meshkov_validity_1965,glick_validity_1965}.
It reduces the dynamics in many-body systems
 to one degenerated band of occupied levels and another degenerated
band of unoccupied levels modeling the typical energy separation of
the HOMO-LUMO gap in closed shell systems. A two-body interaction is
added which generates one prominent coherent resonance excitation.
Depending on the interaction strength, one can simulate a variety of
many-body effects as, e.g., spontaneous symmetry breaking or
large-amplitude collective motion and it has been used for this
purpose particularly in nuclear physics \cite{Hol73a}. The LMG model
is closely related to models of coupled spin-1/2 systems as 
used in quantum optics \cite{Jae06aB}.
The difference lies mainly in the
shaping of interaction which ranges all over the system in the LMG
model while nearest neighbor coupling is often used in other
realizations.  
The LMG model has also been used as a test model in a nuclear context~\cite{Sev06,Lac12,Yil14,Lac16}.
It can be modified to allow for a
description of dissipation by allowing a certain spread of excitation
energies over the levels \cite{reinhard_dissipation_1988}.  The
energies of the levels are distributed stochastically and that is why
we called this extension a Stochastic Two-Level Model (STLM). The
simplicity of the model allows an exact solution and so we use 
STLM here for testing STDHF.

The paper is organized as follows. In Sec.~\ref{sec:theory}, we first present the 
the STLM, then the TDHF, the STDHF and the exact propagation thereof. We also
detail the initial excitation used and the observables studied in this work. And we finally end
the theory section on numerical technicalities. In Sec.~\ref{sec:results}, we discuss the corresponding
results~: we start with a typical test case and then study the impact of varying the model parameters 
and the initial excitation. We finally give some conclusions and perspectives in Sec.~\ref{sec:conclusion}.

\section{Theoretical framework}
\label{sec:theory}
\subsection{A stochastic two-level model}

\begin{figure}[htbp]
\begin{center}
\includegraphics[width=0.8\linewidth]{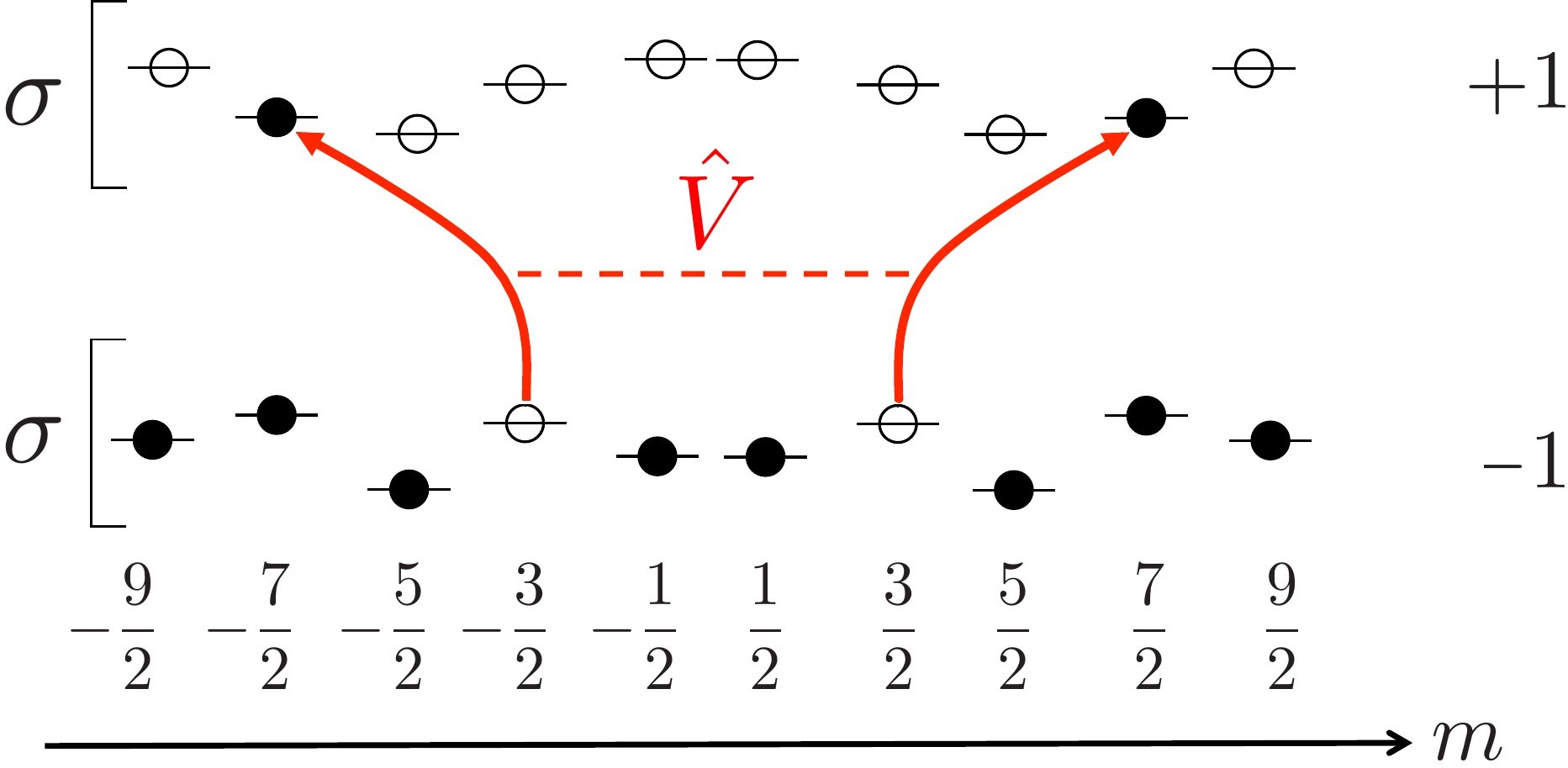}
\caption{\label{fig:Stoch2Level-mod} Illustration of the Stochastic
  Two-Level Model with the action of $\hat{V}$, defined in
  Eq.~(\ref{eq:VS}), applied to the ground state of $\hat{H}_0$,
  defined in Eq.~(\ref{eq:H0}), for $j=9/2$.  }
\end{center}
\end{figure}

The STLM is sketched in Fig.~\ref{fig:Stoch2Level-mod}.
%
The model consists in two bands of single-particle (s.p.) levels, the lower
band denoted by the principle quantum number $s=-1$ and the upper one
by $s=+1$. Each band contains an even number of $\Omega$ levels
denoted by the secondary quantum number $m$ running from $-j$ to $+j$ in
steps of 1 such that  $\Omega=2j+1$ (and $j$ is then half integer).  In the
example displayed in Fig.~\ref{fig:Stoch2Level-mod}, the $j=9/2$ yields a sub-shell with 10 different $m$ values
from $-9/2$ to $+9/2$. 
S.p. states are thus represented by a combined quantum number~:
\begin{subequations}
\begin{equation}
  \alpha=(s_\alpha,m_\alpha)
  \;,\;
  s_\alpha\in\{-1,+1\}
  \;,\;
  m_\alpha\in\{-\!j,-\!j\!+\!1,...,j-1,j\}
  \;.  
\end{equation}
The states are grouped into $\pm m_\alpha$ partners and we keep this
symmetry to reduce the complexity.  For compact notation, we introduce
the following abbreviations wherever convenient
\begin{equation}
  \bar{m}_\alpha=-m_\alpha
  \;,\;
  \bar{\alpha}
  =
  (s_\alpha,\bar{m}_\alpha)
  =
  (s_\alpha,-{m}_\alpha)
  \equiv
  -\alpha
  \;.
\label{eq:alpha}
\end{equation}
\end{subequations}
The notion $\alpha>0$ then means $m_\alpha>0$.

The model Hamiltonian consists out of one-body Hamiltonian $\hat{H}_0$
plus two-body interaction $\hat{V}$. It is constructed in a standard manner on the basis 
of annihilation  (and creation) operators  $\hat{a}_{s_\alpha,{m_\alpha}}^{(\dagger)}=\hat{a}_\alpha^{(\dagger)}$ 
for each s. p. state as~:
\begin{subequations}
\label{eq:Ham1}
\begin{eqnarray}
  \hat{H}
  &=&
  \hat{H}_0+\hat{V}
  \quad, 
\\
  \hat{H}_0
  &=&
  {
  \sum_{\alpha}
  \frac{s_\alpha\varepsilon_\alpha}{2}\hat{a}_{\alpha}^\dagger\hat{a}_{\alpha}^{\mbox {}}
}
\quad,
\label{eq:H0}
\\
  \hat{V}
  &=&
  v_0\hat{S}_{+}\hat{S}_{-}
  \quad,
\label{eq:VS}
\\
  \hat{S}_+
  &=&
  \sum_{\alpha>0}\hat{a}_{\alpha}^\dagger\hat{a}_{\bar{\alpha}}^\dagger
  =
  (\hat{S}_-)^\dagger
  \quad,
\\
  \varepsilon_\alpha
  &=&
  \Delta+\delta \varepsilon_\alpha
  \quad,
\label{eq:defspe}\\
  \delta\varepsilon_\alpha
  &:&
  \sum_{m_\alpha}
    \delta\varepsilon_\alpha \ \stackrel{\Omega \rightarrow +\infty}{\xrightarrow{\hspace*{1cm}}} \ 0
  \quad,\quad
  \frac{1}{\Omega} \sum_{m_\alpha} \left(\delta \varepsilon_\alpha\right)^2
  \ \stackrel{\Omega \rightarrow +\infty}{\xrightarrow{\hspace*{1cm}}} \ 
  \sigma^2
  \quad,
\end{eqnarray}
\end{subequations}
where $\Delta$ stands for the average level spacing between the two shells $ s=-1$ and
$s=+1$ and the $\delta\varepsilon_\alpha$ are chosen
stochastically according to a Gaussian distribution with width
$\sigma$ and centroid zero.  The gap $\Delta$ defines the energy unit
and the time unit is accordingly $[\Delta^{-1}]$. We shall use these units 
all over the text. 
With $\sigma=0$, the model reduces to the case
with fully degenerated bands.  Finally, $v_0$ describes
the strength of the coupling $\hat{V}$.
This interaction is in the form of a pairing interaction as
used in the seniority model or BCS (see sections 6.2 and 6.3 
of \cite{Rin80aB}). It models in most simple manner
collisions between $\alpha$-$\bar{\alpha}$ pairs of fermions.

In the following, we always consider half-filled systems such that the
particle number becomes $N=\Omega$.  In addition, we only consider
weak and repulsive interactions $v_0>0$. This minimizes the effect of
$\hat{V}$ on the ground state such that $\hat{V}$ serves mainly to
induce correlations. From a more physical perspective, it also mocks up 
typical systems in which the mean-field 
provides a good description of ground state properties. This is also well suited for our purpose to
study the treatment of dynamical correlations with STDHF.  The ground
state $|\Phi_0\rangle$ of the free Hamiltonian $\hat{H}_0$ possesses
all electrons in the lower band $s=-1$.  This feature is still
approximately correct for the weak interactions $\hat{V}$ considered
here and it remains even exact at Hartree-Fock (HF) level.

\subsection{The Hartree-Fock (HF) approach}
\label{sec:mf}

The operator $\hat{a}_{\alpha}^\dagger$ creates a single particle
(s.p.) basis state and $\hat{a}_{\alpha}^{\mbox{}}$ annihilates it.
The creation operator
for any other s.p. state is obtained by the linear combination
\begin{equation}
\label{eq:bk}
  \hat{b}^\dagger_\kappa
  =
  \sum_{\alpha>0}\hat{a}^\dagger_\alpha A_{\alpha\kappa}
\quad , \quad
  \hat{b}^\dagger_{\bar{\kappa}}
  =
  \sum_{\alpha>0}\hat{a}^\dagger_{\bar{\alpha}}
  A_{\bar{\alpha}\bar{\kappa}} \quad ,
\end{equation}
where $\kappa>0$ and $\bar{\kappa}=-\kappa$.  The symmetry of $\hat{H}$ allows us to
skip the cross couplings $\alpha\leftrightarrow\bar{\kappa}$ and
$\bar{\alpha}\leftrightarrow{\kappa}$.  A general independent-particle
state (Slater state) for $N=\Omega$ particles is generated by applying
all $\hat{b}^\dagger_\kappa$ ~:
\begin{equation}
  |\Phi\rangle
  =
  \hat{b}_{\kappa_1}^\dagger
  \hat{b}_{\kappa_2}^\dagger \ldots \hat{b}_{\kappa_{\Omega}}^\dagger
  |\mbox{vac}\rangle
  \quad.
  \label{eq:kapbasis}
\end{equation}
The energy expectation value of this state is the Hartree-Fock energy
\begin{subequations}
\begin{eqnarray}
  E_\mathrm{HF}
  &=&
  \langle\Phi|\hat{H}|\Phi\rangle
\nonumber\\
  &=&
  \frac{1}{2}
  \sum_{m_\alpha>0}\varepsilon_{m_\alpha}
  \left(
  \rho_{1m_\alpha,1m_\alpha}-\rho_{-1m_\alpha,-1m_\alpha}
  \right)
  +
  v_0\sum_{\alpha,\alpha'>0}
  \rho_{\alpha'\alpha}\rho_{\bar{\alpha}'\bar{\alpha}}
  \quad,
\\
  \rho_{\alpha'\alpha}
  &=&
  \langle\Phi|\hat{a}^\dagger_\alpha\hat{a}^{\mbox{}}_{\alpha'}|\Phi\rangle
  =
  \sum_{\kappa>0} n_\kappa A^*_{\alpha\kappa}A^{\mbox{}}_{\alpha' \kappa}
  \quad,
\\
  \rho_{\bar{\alpha}'\bar{\alpha}}
  &=&
  \langle\Phi|\hat{a}^\dagger_{\bar{\alpha}}\hat{a}^{\mbox{}}_{\bar{\alpha}'}|\Phi\rangle
  =
  \sum_{\kappa>0} n_{\bar{\kappa}} A^*_{\bar{\alpha}\bar{\kappa}}A^{\mbox{}}_{\bar{\alpha}' \bar{\kappa}}
\quad,
\end{eqnarray}
\end{subequations}
where $\rho_{{\alpha}' {\alpha}}$ is the one-body density matrix and
$n_\kappa$ is the occupation number of state $\kappa$.  The ground
state of the mean-field approximation is obtained by minimizing
$E_\mathrm{HF}$ with respect to the $\rho_{\alpha'\alpha}$ or to
$A^{\mbox{}}_{\alpha'\kappa}$. For the regime of weak $v_0>0$ which
we are studying, the mean-field ground state $|\Phi_0\rangle$ of the
interacting system is identical to the ground state of $\hat{H}_0$
which is given by the trivial non-transformation
$A^{\mbox{}}_{\alpha\kappa}=\delta_{\alpha\kappa}$.

%
The TDHF equations are derived by the time-dependent variational
principle \cite{Mar10aB} and solved in practice by expressing them in
terms of the amplitudes $A^{\mbox{}}_{\alpha\kappa}(t)$, yielding
\begin{equation}
  \mathrm{i}\hbar\partial_tA^{\mbox{}}_{\alpha\kappa}
  =
  \sum_{\alpha'}\hat{h}_{\alpha\alpha'}A^{\mbox{}}_{\alpha'\kappa}
  \quad,\quad 
\hat{h}_{\alpha \alpha'} = \frac{\varepsilon_{m_\alpha}}{2} s_\alpha \delta_{\alpha \alpha'}
+ \rho_{\bar{\alpha}' \bar{\alpha}}
\quad .
\end{equation}
The numerical solution is done using an implementation of the
Crank-Nicolson scheme \cite{crank_practical_1947}
\begin{equation}
  \hat{A}(t\!+\!dt)
  =
  \frac{1-\frac{i \textrm dt}{2\hbar}\hat{h}(t+\textrm dt/2)}
       {1+\frac{i \textrm dt}{2\hbar}\hat{h}(t + \textrm dt/2)}\hat{A}(t)
  \quad,
\label{eq:CN-HF}
\end{equation}
where $\hat{A}$ is a compact notation of the matrix $A_{\alpha\kappa}$
of expansion coefficients. $\hat{h}(t +\textrm dt/2)$ is computed in a
predictor step which looks like the full step (\ref{eq:CN-HF}) but
propagating only by $\textrm dt/2$ and using $\hat{h}(t)$.  The
Crank-Nicolson step maintains ortho-normality by construction. 
To obtain satisfying energy conservation, one has to choose the
step size $\textrm dt$ sufficiently small.

\subsection{Stochastic Time-Dependent Hartree Fock (STDHF)}

Mean-field propagation with TDHF, as outlined in Sec.~\ref{sec:mf},
takes only part of the two-body interaction $\hat{V}$ into account.
There remains a residual interaction from $\hat{V}$ which generates
correlations. The idea in STDHF is to simplify the description of
those correlations by expressing a correlated state as an (incoherent)
ensemble of mean-field states and the propagation of correlations by
occasional two-particle-two-hole ($2p2h$) jumps
\cite{reinhard_stochastic_1992,Sur14a}, which makes sense in 
dynamical regimes attained at sufficiently high excitation energies, when exploring 
dense enough excitation spectra. The idea is somewhat similar to the ideas underlying 
the derivation of the 
semi-classical Vlasov-Uehling-Uhlenbeck (VUU) approach \cite{Ber88,Abe96R}, where
dynamical correlations are reduced to incoherent two-particle
collisions. It can even be formally shown that STDHF, once properly averaged, reduces to 
the stochastic  Boltzmann-Langevin equation \cite{reinhard_stochastic_1992}.

The state of the system is described by an ensemble of Slater states
\begin{equation}
  \{|\Phi^{(i)}(t)\rangle,i=1,\ldots,\mathcal{N}_\mathrm{ens}\}
  \quad.
\end{equation}
Although not computed explicitly, we assume that each
state is allowed to develop $2p2h$ correlations as~:
\begin{subequations}
\begin{eqnarray}
  |\Phi^{(i)}\rangle
  \longrightarrow
  |\Psi^{(i)}\rangle
  &=&
  c_0^{(i)}|\Phi^{(i)}\rangle
  +
  \sum_{\kappa_1\kappa_2\kappa_3\kappa_4}
  c^{(i)}_{\kappa_1\kappa_2\kappa_3\kappa_4}
  |\Phi^{(i)}_{\kappa_1\kappa_2\kappa_3\kappa_4}\rangle
  \;,
\\
  |\Phi^{(i)}_{\kappa_1\kappa_2\kappa_3\kappa_4}\rangle
  &=&
  (\hat{b}^{(i)})^\dagger_{\kappa_1}
  (\hat{b}^{(i)})^\dagger_{\kappa_2}\,
  \hat{b}^{(i)}_{\kappa_3}\,
  \hat{b}^{(i)}_{\kappa_4}
  |\Phi^{(i)}\rangle
  \;,
\label{eq:correl2ph}
\end{eqnarray}
\end{subequations}
where $\kappa_1$, $\kappa_2$ are unoccupied states
and $\kappa_3$, $\kappa_4$ occupied ones. 
For simplicity, we did not write
explicitly the time argument in the wave functions and operators.
After a certain time of propagation $\tau_\mathrm{sample}$, the
coherent state (\ref{eq:correl2ph}) is decomposed into an incoherent
ensemble of mean-field states, which are attainable in a probabilistic manner after 
this incoherent reduction.  We can then evaluate transition probabilities  in time-dependent
perturbation theory. This yields the transition probabilities for a
jump from the Slater state $|\Phi^{(i)}\rangle$ to the Slater state
$|\Phi^{(i)}_{\kappa_1\kappa_2\kappa_3\kappa_4}\rangle$ as
\begin{subequations}
\begin{eqnarray}
\MD{
  w_{\kappa_1\kappa_2\kappa_3\kappa_4}}
  &=&
  \MD{\mathcal{P}_{\kappa_1\kappa_2\kappa_3\kappa_4} \ \tau_{\rm sample}} 
\label{eq:w_stdhf}\\
  \mathcal{P}_{\kappa_1\kappa_2\kappa_3\kappa_4} &=&
  \frac{2\pi}{\hbar}\ 
  \delta_\Gamma(
  E^{\rm HF}_{\kappa_1\kappa_2\kappa_3\kappa_4}-E^{\rm HF}_0
  )\ 
  \left|\langle\Phi^{(i)}_{\kappa_1\kappa_2\kappa_3\kappa_4}|
  \hat{V}_\mathrm{res}
  |\Phi^{(i)}\rangle\right|^2
\label{eq:P_stdhf}
\end{eqnarray}
\end{subequations}
where $\delta_\Gamma$ is a $\delta$ function with finite width
$\Gamma$. \MD{Detailed studies on the chosen numerical values 
of $\tau_{\rm sample}$ and $\Gamma$ are discussed
in Secs.~\ref{sec:tau} and \ref{sec:Gamma} respectively (see also \cite{Hov55,Sla15a} for more details).}
$E^{\rm HF}_{\kappa_1\kappa_2\kappa_3\kappa_4}$ and $E^{\rm HF}_0$ are
the Hartree-Fock energies of the $2p2h$ and the original state
respectively. It turns out that the sole matrix elements of the
residual interaction which are non-vanishing for the STLM read~:
\begin{equation}
  \langle\Phi^{(i)}_{\kappa_1\bar{\kappa}_2\bar{\kappa}_3\kappa_4}|
   \hat{V}_\mathrm{res}|\Phi^{(i)}\rangle
  =
  v_0
  \sum_{\alpha,\alpha'>0}
  A_{\alpha\kappa_1}^{(i)*}A_{\bar{\alpha}\bar{\kappa}_2}^{(i)*}
  A_{\bar{\alpha}'\bar{\kappa}_3}^{(i)}A_{\alpha'\kappa_4}^{(i)}
\quad,
\end{equation}
where all $\kappa$'s entering the latter equation are now positive.
The decision to jump to
$|\Phi^{(i)}_{\kappa_1\bar{\kappa}_2\bar{\kappa}_3\kappa_4}\rangle$ or
to remain in the original state is done in Monte-Carlo fashion
according to the probability
$w_{\kappa_1\bar{\kappa}_2\bar{\kappa}_3\kappa_4}$.  This is
performed for each $|\Phi^{(i)}\rangle$ in the ensemble.  The ensemble
starts initially from the same state for all $i$.  Then, each
trajectory $i$ develops its own dynamics through the stochastic
choices described above. Finally, observables are computed for an
ensemble average.

\subsection{Exact propagation}
\label{sec:exact}

The exact solution is conceptually the simplest but computationally
most expensive. The fully correlated state is expanded into a
complete basis of mean-field states
\begin{equation}
  |\Psi\rangle
  =
  \sum_{\alpha_n,\alpha'_n>0}
  c_{\alpha_1...\alpha_{\Omega/2},{\bar{\alpha}'}_1...{\bar{\alpha}'}_{\Omega/2}}
  \hat{a}^\dagger_{\alpha_1}...\hat{a}^\dagger_{\alpha_{\Omega/2}}
  \hat{a}^\dagger_{{\bar{\alpha}'}_1}...\hat{a}^\dagger_{{\bar{\alpha}'}_{\Omega/2}}
  |\mbox{vac}\rangle
  \;.
  \label{eq:expan}
\end{equation}
The time-dependent Schr\"odinger equation
\begin{equation}
  \mathrm{i}\hbar{\partial_t}|\Psi\rangle
  =
  \hat{H}|\Psi\rangle
\end{equation}
is solved by mapping it into a matrix equation for the expansion
coefficients
$c_{\alpha_1...\alpha_{\Omega/2},{\bar{\alpha}'}_1...{\bar{\alpha}'}_{\Omega/2}}$.
For the solution, we use again the Crank-Nicolson scheme
(\ref{eq:CN-HF}), but now with the full Hamiltonian
$\hat{H}$.  The inverse $(1+\frac{\mathrm{i} \textrm dt}{2\hbar}\hat{H})^{-1}$
appearing therein is computed by solving a linear system using a
bi-conjugate gradient stabilized method \cite{vorst_bi-cgstab:_2006}.

\subsection{Initial excitation}
\label{sec:init}

As a first step, we have to prepare the ground state of the
stationary problem.  This is for TDHF and STDHF the Slater state with
each $s_\alpha=-1$ s.p. state occupied and each $s_\alpha=+1$
unoccupied. For the exact solution, we have  to solve the static
Schr\"odinger equation with the full Hamiltonian $\hat{H}$. 

The initial state for dynamical evolution is then obtained from the ground
state by an instantaneous boost excitation
\begin{subequations}
\label{eq:init_excit}
\begin{eqnarray}
  |\Phi(t\!=\!0)\rangle
  &=&
  e^{\mathrm{i}\lambda(\hat{D}+\gamma \hat{W})}|\Phi_\mathrm{gs}\rangle
  \quad,
\\
  \hat{D}
  &=&
  \sum_{m_\alpha}
\left(
  \hat{a}_{1\,m_\alpha}^\dagger\hat{a}_{-1\,m_\alpha}^{\mbox {}} 
  +
  \hat{a}_{-1\,m_\alpha}^\dagger\hat{a}_{1\,m_\alpha}^{\mbox {}}
\right)
  \quad,
\label{eq:dipoleOp}
\\
  \hat{W}
  &=&
  {\frac{1}{2}\sum_{\alpha,\alpha'>0}
  \left(
\hat{a}_{\alpha}^\dagger \hat{a}_{\alpha'}^{\mbox {}} 
  +
  \hat{a}_{\bar{\alpha}}^\dagger \hat{a}_{\bar{\alpha}'}^{\mbox {}}
\right)}
  \quad.
\end{eqnarray}
\end{subequations}
The $\hat{D}$ simulates a dipole operator of a typical many-particle
system and the excitation operator $e^{\mathrm{i}\lambda\hat{D}}$ 
induces initial $1p1h$ transitions within the same $m_\alpha$. For example, one can consider the operator that couples
a laser field to the electrons of an atom, and an instantaneous boost
is the most generic excitation simulating a short pulse. 
The parameter $\lambda$ tunes the strength of the initial excitation.

The $\hat{W}$ operator serves a different purpose. 
At the mean-field level,
the interaction (\ref{eq:VS}) 
deals only with vertical transitions which maintain the $m$
quantum number. This emphasizes coherence which, in turn, overlays
dissipation with large memory effects as we will see. To explore the
level of dissipation in a more flexible manner, we stir up the
interaction with transition across different $m$'s by applying the unitary transformation
$e^{\mathrm{i}\lambda\gamma\hat{W}}\hat{V}e^{-\mathrm{i}\lambda\gamma \hat{W}}$ on the coupling
$\hat{V}$ with the
mixing operator $\hat{W}$. This transformation maintains the overall interaction strength without rescaling
$v_0$. For simplicity, we apply the transformation to the initial
state which turns out to be a good approximation to solving the
dynamical equations with the modified interaction.
The parameter $\gamma$ quantifies the amount
of mixing with respect to the dipole operator.

\subsection{Observables}
\label{sec:observ}

Our main aim here is to study the dynamics of thermalization after initial
excitation. This is quantified by the fermionic entropy
\begin{equation}
  S
  =
  -\Tr \left[\hat{\rho}\ln\hat{\rho}+(1-\hat{\rho})\ln(1-\hat{\rho}) \right] \ .
\label{eq:St}
\end{equation}
where $\hat{\rho}$ is the one-body density operator whose matrix
elements are the one-body density matrix
$\rho_{\alpha'\alpha}=\langle\hat{a}^\dagger_\alpha\hat{a}^{\mbox{}}_{\alpha'}\rangle$.
A pure mean field state has $S=0$. The entropy thus stays zero at all
time in any TDHF calculation, while STDHF and the exact solution are
expected to exhibit a time-dependent $S$.

A further test observable is the difference between one-body matrices
in all combinations between TDHF, STDHF and the exact solution,
quantified as
\begin{equation}
  \delta_\rho
  = 
  2\frac{||\rho-\rho_{\rm ex}||}{||\rho+\rho_{\rm ex}||} \quad,
\label{eq:diffnorm}
\end{equation}
where $||...||$ stands for the Frobenius matrix norm
$||A||=\sqrt{\Tr(\hat{A}^\dagger\hat{A})}$.
 
\subsection{\MD{Model parameters}}
\label{sec:num}

We complete this section with specifying the values of the model parameters.  To stay at the
perturbative level for the correlations, the two-body interaction is
taken relatively small, that is $v_0=0.05\, \Delta$.  We
propagate TDHF and the exact solution using the Crank-Nicolson scheme
with a time step $\mathrm{d}t=0.01\,\Delta^{-1}$.  

The number of particles varies from $\Omega=4$ to 10 in the comparison
between the exact solution and the STDHF. Due to an exponentially
increasing computing time for the exact propagation, the cases from
$\Omega=12$ to $20$ have been explored in the STDHF scheme only.  The
number of samples in the STDHF ensemble is always $\mathcal
N_\mathrm{ens}=100$. We have checked also larger ensembles and found
practically the same results.

\begin{figure}[htbp]
\begin{center}
\includegraphics[width=0.7\linewidth]{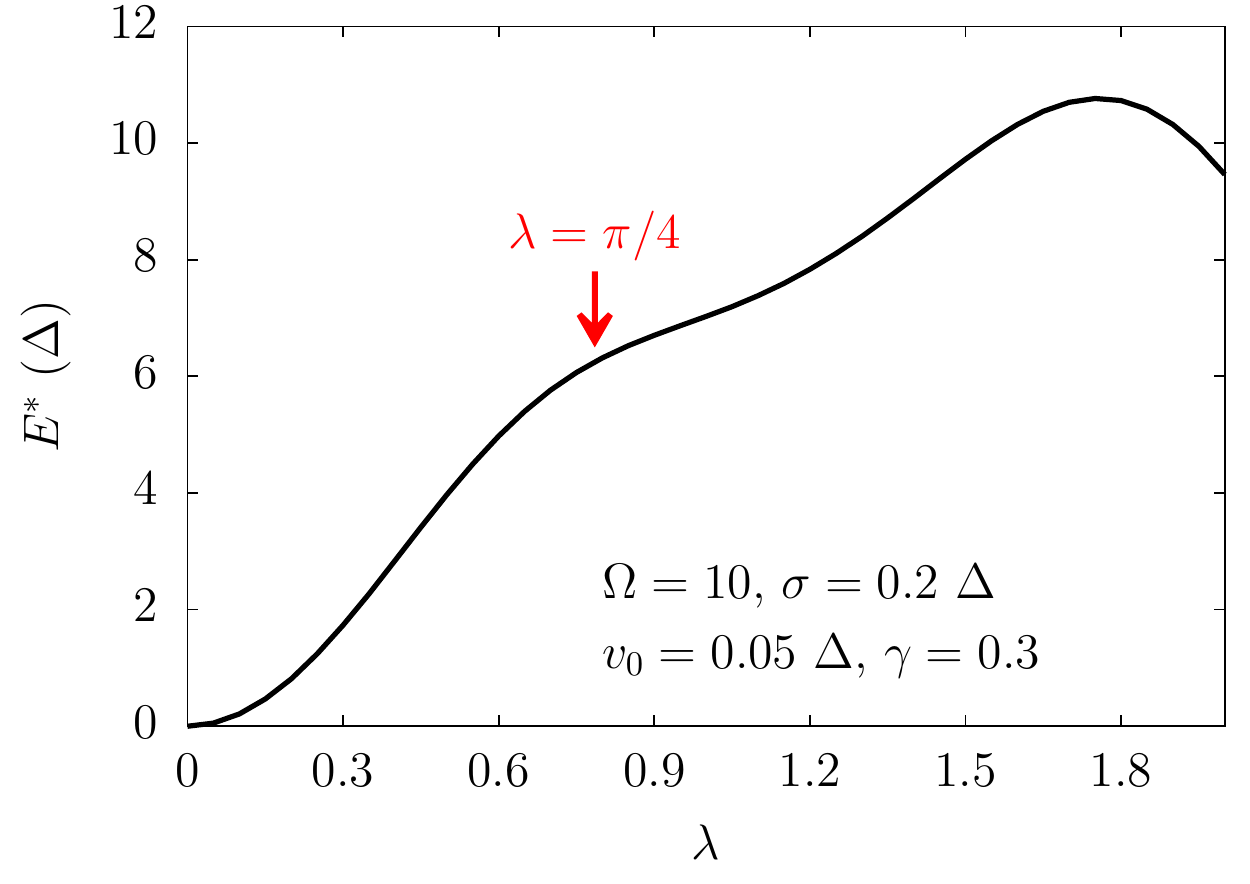}
\caption{
Accessible excitation energy $E^*$ (in $\Delta$ units) as a function
of $\lambda$ for the initial excitation (\ref{eq:init_excit}) with $\gamma=0.3$, and the other 
STLM parameters as indicated.
\label{fig:init_excit}
}
\end{center}
\end{figure}

\MD{We have tested various values of the mixing parameter $\gamma$, from 0 to 0.6.
In the following results, it is set to 0.3 since such a value creates 
enough disorder at $t=0$, that is the needed transitions for STDHF to operate (see 
discussion in Sec.~\ref{sec:init}).}
Figure~\ref{fig:init_excit} shows the excitation energy $E^*$ attained
as a function of $\lambda$ in the case of 10 particles, a band width $\sigma=0.2~\Delta$ 
and an interaction strength $v_0 = 0.05~\Delta$.
The energy grows
monotonously up to a maximum where it turns to monotonous decrease.
The upper limit of $E^*$ reflects the fact that the model Hamiltonian
is bound not only from below, as it should be for a non-relativistic
Hamiltonian, but also from above. The relevant range of the STLM
stays in the region of $\lambda$ safely below the turning point (which is at
$\lambda=\pi/2$ for $\gamma=0$). In practice, we use $0.6 \leq \lambda \leq 0.8$ close to $\pi/4$
which corresponds to a state with the particles having half weight in the upper band and half
weight in the lower band.  This value for $\lambda$ is safely in the
regime of increasing $E^*$.
We have checked that the cases $\gamma=0$ and $\gamma=0.3$ provide moderate differences
in $E^*$. This corroborates the above statement that
the overall interaction strength is little affected by virtue
of scanning interactions in terms of a unitary transformation.


\section{Results}
\label{sec:results}

\subsection{A first test case}
\label{sec:test_case}
\begin{figure}[htbp]
\includegraphics[width=0.8\linewidth]{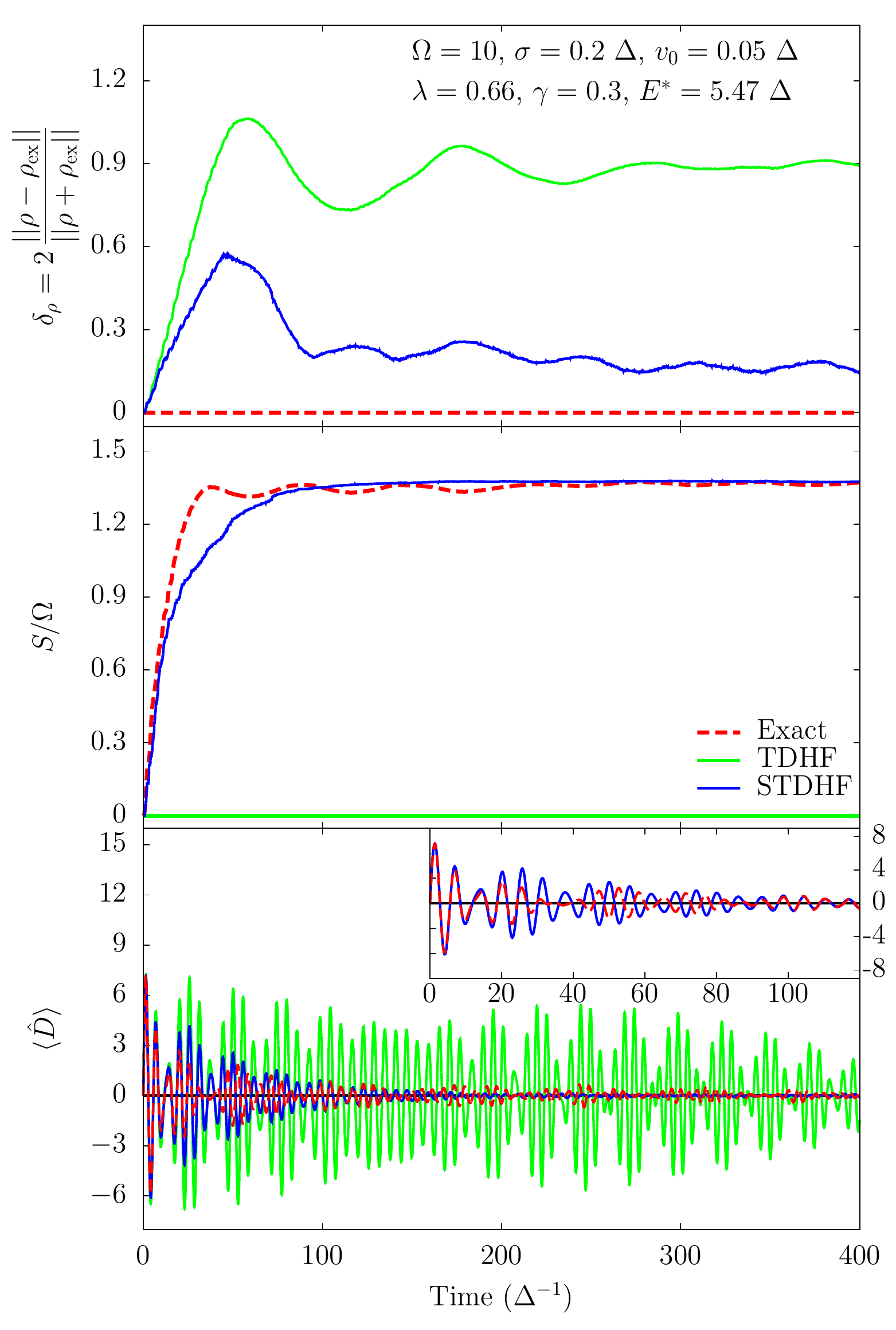}
\caption{\label{fig:test_case} Comparison of TDHF, STDHF and exact
solution with respect to three observables, as a function of time, with the STLM parameters
as indicated.
Bottom~: expectation value of the dipole operator $\hat{D}$,
see Eq.~(\ref{eq:dipoleOp}), \MD{with as an insert a zoom on the exact and the STDHF
responses below $120~\Delta^{-1}$}. Middle~: entropy per particle $S/\Omega$, see 
Eq.~(\ref{eq:St}). Top~: difference of one-body density.
}
\end{figure}

We start with the analysis of a typical test case from the perspective
of the difference of s.p. density matrices, the entropy, and the expectation value of the dipole.  
The time evolution of these
observables is displayed in Fig.~\ref{fig:test_case}.
In the upper panel, we compare the density matrix $\rho$ obtained in
HF and that in STDHF with respect to the exact density matrix
$\rho_{\rm ex}$ in terms of norm of the difference $\delta_\rho$, defined in 
Eq.~(\ref{eq:diffnorm}), of TDHF or STDHF with respect to the exact
solution.  STDHF provides a much smaller deviation $\delta_\rho$ than TDHF.
This indicates that STDHF is indeed able to incorporate a great deal
of the dynamical correlations. Both deviations are composed from a
global trend plus oscillations. The trend approaches a rather large
stable deviation $\delta_\rho$ for TDHF and seems to indicate (slow)
convergence to negligible deviation for STDHF. 

These oscillations are also observed in the time evolution of the
exact entropy per particle, see middle panel of
Fig.~\ref{fig:test_case}. 
\PGR{The entropy from STDHF entropy reproduces the exact entropy 
in the average trend and in the asymptotic value rather well, see also
Figure~\ref{fig:entropy_sigma} and the related discussion.} 
The main discrepancies lie in the lack of oscillations. Indeed, the instantaneous
(Markovian) approximation when evaluating the stochastic jumps in
STDHF erases at once all coherence and memory effects.
Similar features are  seen in semi-classical VUU models where the collision
term is also treated in Markovian approximation. The oscillating
entropy for the exact solution shows that the STLM 
still carries a substantial amount
of memory effects which can only be coped with using a frequency (or
time) dependent kernel for the jump probability \cite{Gre94a}.  
\PGR{The TDHF result (light green line) differs dramatically from STDHF.
It maintains zero entropy throughout and
never relaxes to anything like a thermalized state. It cannot
reproduce at all the long-time behavior of system as soon as
dissipation becomes relevant.}

\PGR{
It is also interesting to note that the maximum value of entropy $S/N=2\log 2 \simeq1.38$
corresponds to an equidistribution of $N=10$ particles
over all $2\Omega=20$ states with occupation probability
$w_\alpha=\rho_{\alpha\alpha}=1/2$. To that extent, the agreement at
the maximum is sort of trivial. However, the STDHF results follow the
exact curve also down to lower values and agree as long as the STDHF
jumps have their grip. This is the non-trivial result indicating that
STDHF catches the basic statistical properties of the system.
}

The lower panel
of Fig.~\ref{fig:test_case} shows the evolution of
dipole momentum for the three solution schemes. 
This confirms what we had seen already from the two other
observables. \PGR{TDHF shows long standing oscillations and
reverberations and thus stays far off the exact solution. 
STDHF, on the other hand, constitutes a remarkable improvement,
in particular over the first $120~\Delta^{-1}$.
STDHF is thus providing
a reliable description of dissipation. A slight difference comes up at
later times. The exact solution shows some reverberations which are
absent in STDHF. These reverberations are related to the oscillations
in entropy and thus an effect of quantum coherence deliberately
suppressed in  STDHF. But this is a small effect and the generally
good agreement prevails. 
} 

This first example shows strengths and weaknesses of STDHF. It is a
therefore a great improvement as compared to TDHF in that it properly catches the
dissipative aspects of many-dynamics and produces in due time the
correct asymptotic state (thermal equilibrium). However, STDHF implies
a Markovian approximation (that is, instantaneous jumps) and is unable to
incorporate memory effects. It then depends on the system and
dynamical regime how important memory effects are. 

\PGR{The three observables shown in Fig.~\ref{fig:test_case} are of
  different nature and show different aspects of the system. Dipole
  momentum $\langle\hat{D}\rangle$ and one-body entropy $S$ are both
  one-body observables where $S$ characterizes the state of the system
  while $\langle\hat{D}\rangle$ demonstrates measurable consequences
  of dissipation. The information concerning dissipation is
  comparable.  We prefer in the following the entropy $S$ because it
  displays the simpler signal. 

The norm of the difference of full
density matrices, $\delta_\rho$, is a many-body observable and so
also carries many-body information not accounted for in comparing
one-body entropies. Still,
$\delta_\rho$ and $S$ deliver comparable information to the extent
that STDHF is much closer to the exact solution than TDHF. There is a
faint difference too: $\delta_\rho$ is more critical as it shows
always a small, but finite, value for STDHF, indicating the known fact
that STDHF unavoidably sacrifices parts of the exact solution, namely
the coherent correlations. We are interested here in an appropriate
reproduction of dissipation which is biased in incoherent
correlations. The one-body entropy $S$ is the more appropriate measure
for this purpose.  We thus confine the further examples to the entropy
only.
}

%

\MD{
\subsection{Impact of the sampling time}
\label{sec:tau}

Two important ingredients of the STDHF transition probability given in Eq.~(\ref{eq:P_stdhf})
are the sampling time $\tau_{\rm sample}$ and the finite width of the $\delta_\Gamma$ function, see~\cite{Sla15a} for
a detailed discussion. We here specifically explore the impact of the first one in the studied model.
Indeed, $\tau_\mathrm{sample}$ has to be large enough such that the
oscillations of the wave functions' phase satisfy the $\delta_\Gamma$ function.
On the other hand, it should be small enough to 
resolve the temporal changes of the mean-field. In addition, it should remain smaller than the total jump rate, i.e.
\begin{equation}
  \tau_{\rm sample} 
  \leq \left(
  \sum_{\kappa_1,\kappa_2,\kappa_3,\kappa_4}
\mathcal{P}_{\kappa_1 \kappa_2 \kappa_3\kappa_4}\right)^{-1}
  \quad.
\label{eq:jump_rate}
\end{equation}

Figure \ref{fig:entropy_tau_t} illustrates
the impact of the variation of $\tau_{\rm sample}$ on the example of the detailed time 
evolution of the one-body entropy. 
\begin{figure}[h]
\centerline{\includegraphics[width=\linewidth]{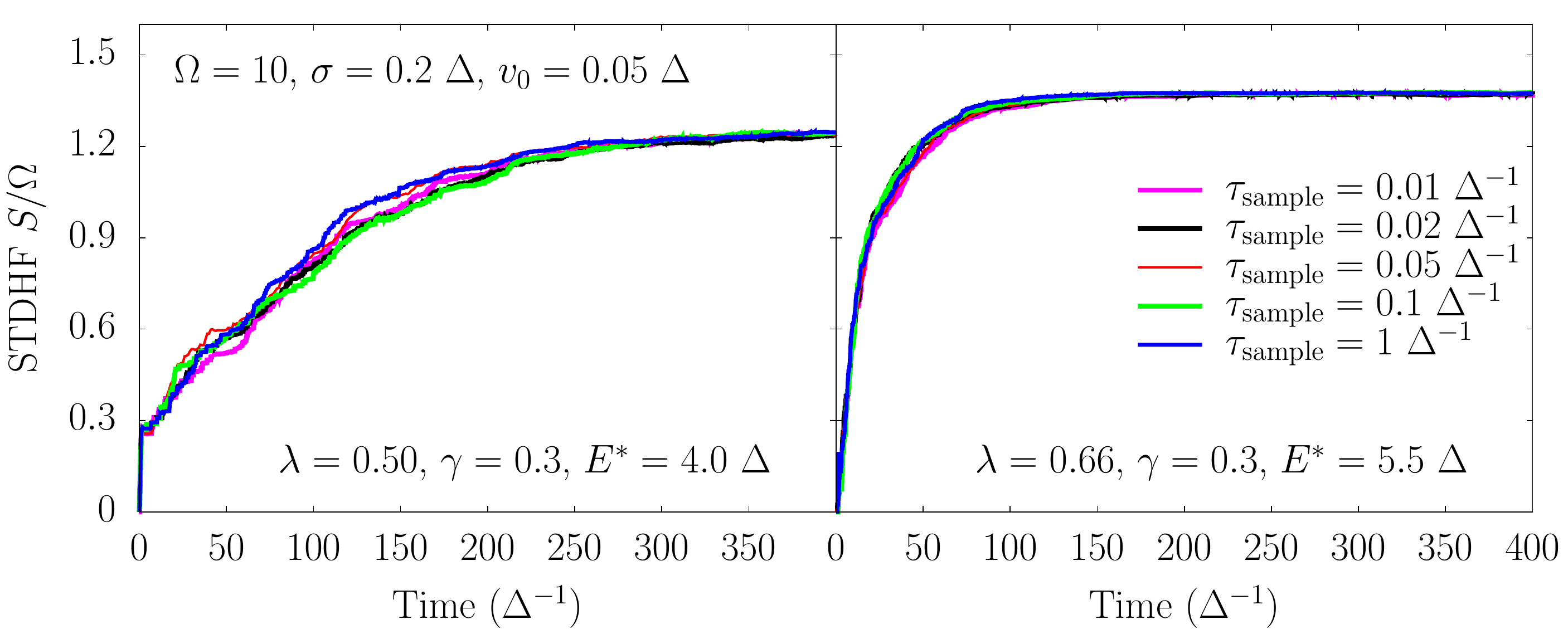}}
\caption{\label{fig:entropy_tau_t}
Time evolution of one-body entropy $S$ for 5 different sampling
times $\tau_\mathrm{sample}$ as indicated, and for two different excitation
energies, $E^*=4.0~\Delta$ (left) and $5.5~\Delta$ (right). }
\end{figure}
$\tau_\mathrm{sample}$ ranges from $0.01~\Delta^{-1}$ to $1~\Delta^{-1}$, that is over 
two orders of magnitude. The corresponding results are stable within
this rather broad parameter window. Note however that, for some values of $E^*$, 
the case $\tau_{\rm sample}=0.1$ or $1~\Delta^{-1}$ exceeds
the total transition rate. Therefore, in all the following results, we have set
$\tau_\mathrm{sample}=5~\textrm dt=0.05~\Delta^{-1}$ to stay on the safe side and to satisfy
 inequality~(\ref{eq:jump_rate}).

\subsection{Impact of the finite width $\Gamma$}
\label{sec:Gamma}

We now discuss the influence of the finite width $\Gamma$ of the $\delta$ function entering
Eq.~(\ref{eq:P_stdhf}).
Unlike semi-classical cases where we deal with a continuum
\cite{Ber88,Abe96R}, such a finite width is necessary in a discrete
quantum spectrum to grab all possibly relevant transitions. 
However, it should be chosen small enough to maintain an acceptable
energy conservation \cite{Hov55,Sla15a}. 
In practice, the $\delta_\Gamma$ function is approximated by a fixed window 
\begin{equation}
 \delta_\Gamma(x)
 =  \left\{\begin{array}{lcl}
  1/\Gamma &\mbox{for}& |x| <\Gamma/2
 \\
  0 &\mbox{for}& |x|\geq \Gamma/2
 \end{array}\right.
 \quad.
\label{eq:box}
\end{equation}

Figure
\ref{fig:entrop_ener_dw_t} demonstrates the effect of $\Gamma$ which has been varied
from $10^{-4}~\Delta$ to $0.1~\Delta$.  
\begin{figure}[h]
\centerline{\includegraphics[width=0.8\linewidth]{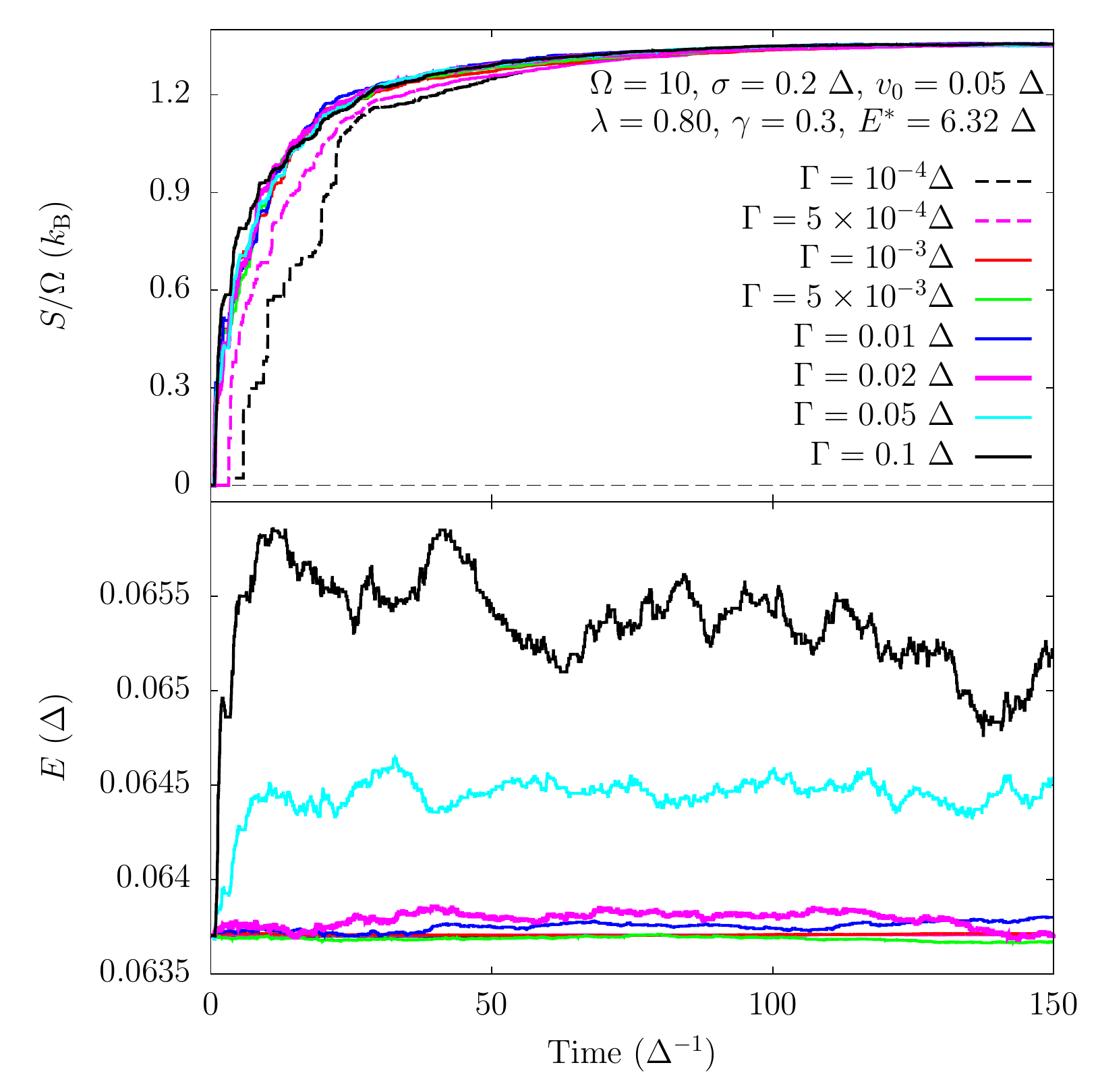}}
\caption{\label{fig:entrop_ener_dw_t}
Time evolution of one-body entropy $S$ (upper panel) and total energy
(lower panel) for different values of the energy width $\Gamma$ in
the energy criterion $\delta_\Gamma$ of the STDHF jump probability, see Eq.~(\ref{eq:P_stdhf}).}
\end{figure}
The smallest values $\Gamma=10^{-4}\Delta$ and $5\times 10^{-4}\Delta$ are
obviously too low and do not produce enough jumps to provide the right rise of the 
one-body entropy $S$ at short times, see upper panel. All higher values yield the same
pattern for $S$ and thus reproduce correctly the dissipation from two-body transitions.
On the other hand, as expected, energy conservation (lower panel) is degraded with increasing
$\Gamma$. Note however the narrow scale for energy in the plot.
Even the worst case still shows a fair energy conservation.  The values
of $\Gamma$ up to $2\times10^{-2}~\Delta$ deliver
a good compromise between appropriate dissipation and energy
conservation. Our chosen value in the following analysis is
$\Gamma=0.02~\Delta$ and constitutes the best compromise in this respect.

We have also checked the effect of the actual profile of the
$\delta_\Gamma$ function by comparison with a Gaussian instead of the
box distribution (\ref{eq:box}).
The differences are marginal. We have therefore preferred 
the box distribution because it is by definition limited and does not have the
danger of occasional outliers as infinite distributions like Gaussians
have. 
}

\subsection{Varying the excitation energy}
\label{sec:alpha}

Dissipation is usually weak in the regime of low excitations and
acquires importance only for sufficiently large excitation energy
\cite{Kad62}. Stochastic evaluation of dynamical correlations as done
in STDHF or VUU is designed for high excitation energies where the
phase space for jumps is (hopefully) dense enough. It is thus of
interest to check the performance with varying excitation energy
\PGR{$E^*=E-E_\mathrm{g.s.}$ which is the difference between the
  actual (conserved) energy $E$ of the system and the ground-state
  energy $E_\mathrm{g.s.}=-\sum_{m_\alpha}\varepsilon_{-1,m_\alpha}$.}

\begin{figure}[htbp]
\begin{center}
\includegraphics[width=0.8\linewidth]{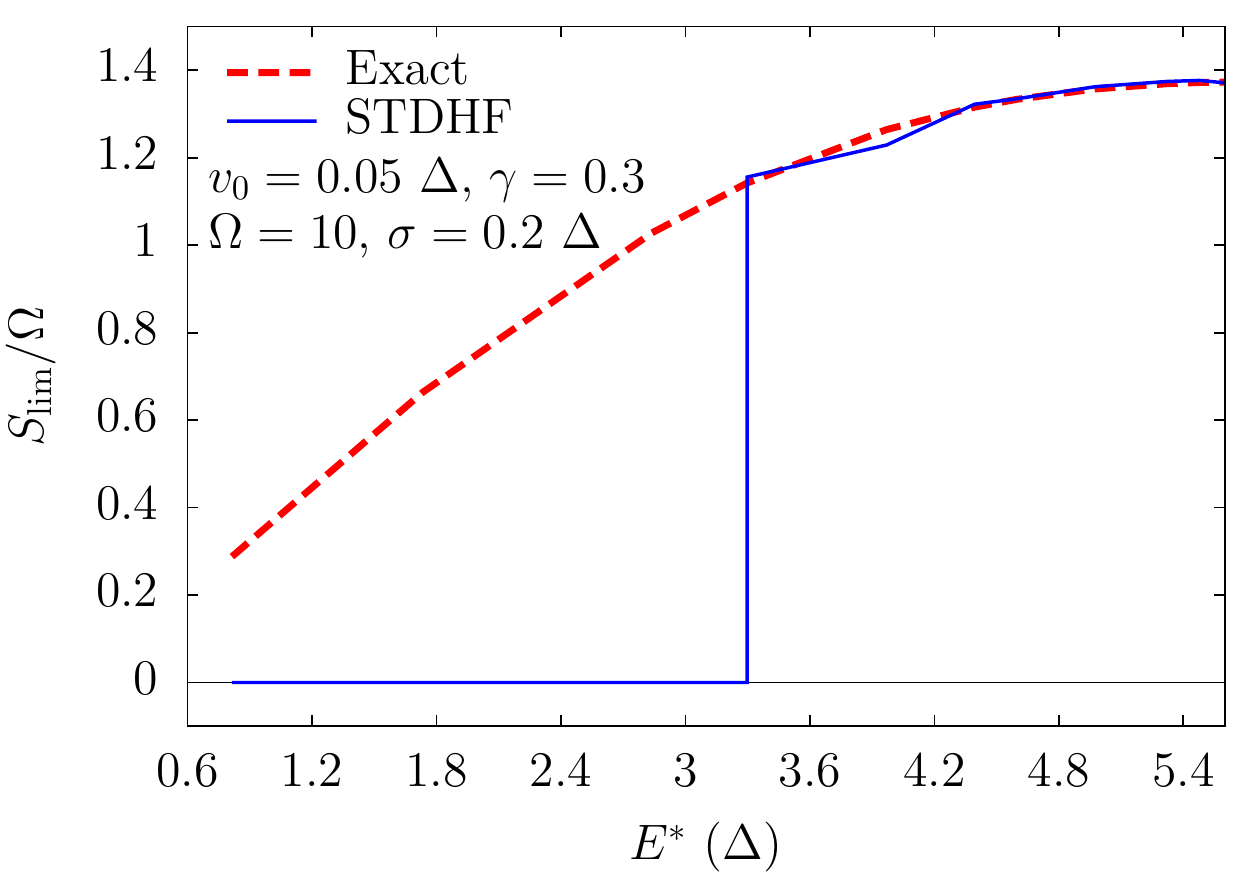}
\caption{\label{fig:s_alpha_estar} Asymptotic entropy per particle
  $S_\mathrm{lim}/\Omega$ for the exact solution (thick curves) and
  for STDHF (thin curves), as a function of the excitation energy $E^*$
  obtained by scan of initialization strength $\lambda$, see
  Eq.~(\ref{eq:init_excit}), and the other
  parameters of the model as indicated.}
\end{center}
\end{figure}

Figure \ref{fig:s_alpha_estar} shows the asymptotic one-body entropy $S_{\rm lim}$
as a function of $E^*$ (in the branch of increasing
excitation energy, see discussion of Fig.~\ref{fig:init_excit}). 
The exact solution shows a smooth and monotonous increase. 
STDHF behaves much different in that it shows a threshold behavior.  It
remains inactive for $E^*<3.3~\Delta$ and suddenly switches to reproduce the
exact value once the stochastic jumps get their grip. 
The result confirms that STDHF is an approach for sufficiently high
excitation energy. 
\MD{We have also explored a profile different from a box function for the finite $\delta_\Gamma$ 
function defined in Eq.~(\ref{eq:box}), namely we have tested a Gaussian instead. The energy
threshold slightly depends on the profile chosen for $\delta_\Gamma$
but the asymptotic value $S_{\rm lim}$ remains unchanged regardless $\delta_\Gamma$
or even the width $\Gamma$ itself.}

\PGR{A comment about maximum value of the entropy shown in 
Fig.~\ref{fig:s_alpha_estar} is in order. It comes very close to the
  value $S/\Omega=2\log{2}\approx 1.38$. As already discussed in Sec.~\ref{sec:test_case},
it stands for
  equi-partition $\rho_{\alpha\alpha}=N/(2\Omega)=1/2$ and is the
  maximally possible value for $S/\Omega$. The figure shows that, in
  the given mode, STDHF comes to work only for rather flat
  distributions of $\rho_{\alpha\alpha}$ with significant occupation
  of the upper band, thus having $S/\Omega$ near the maximum.}

\subsection{Impact of band width}
\label{sec:band}

\begin{figure}[htbp]
\includegraphics[width=\linewidth]{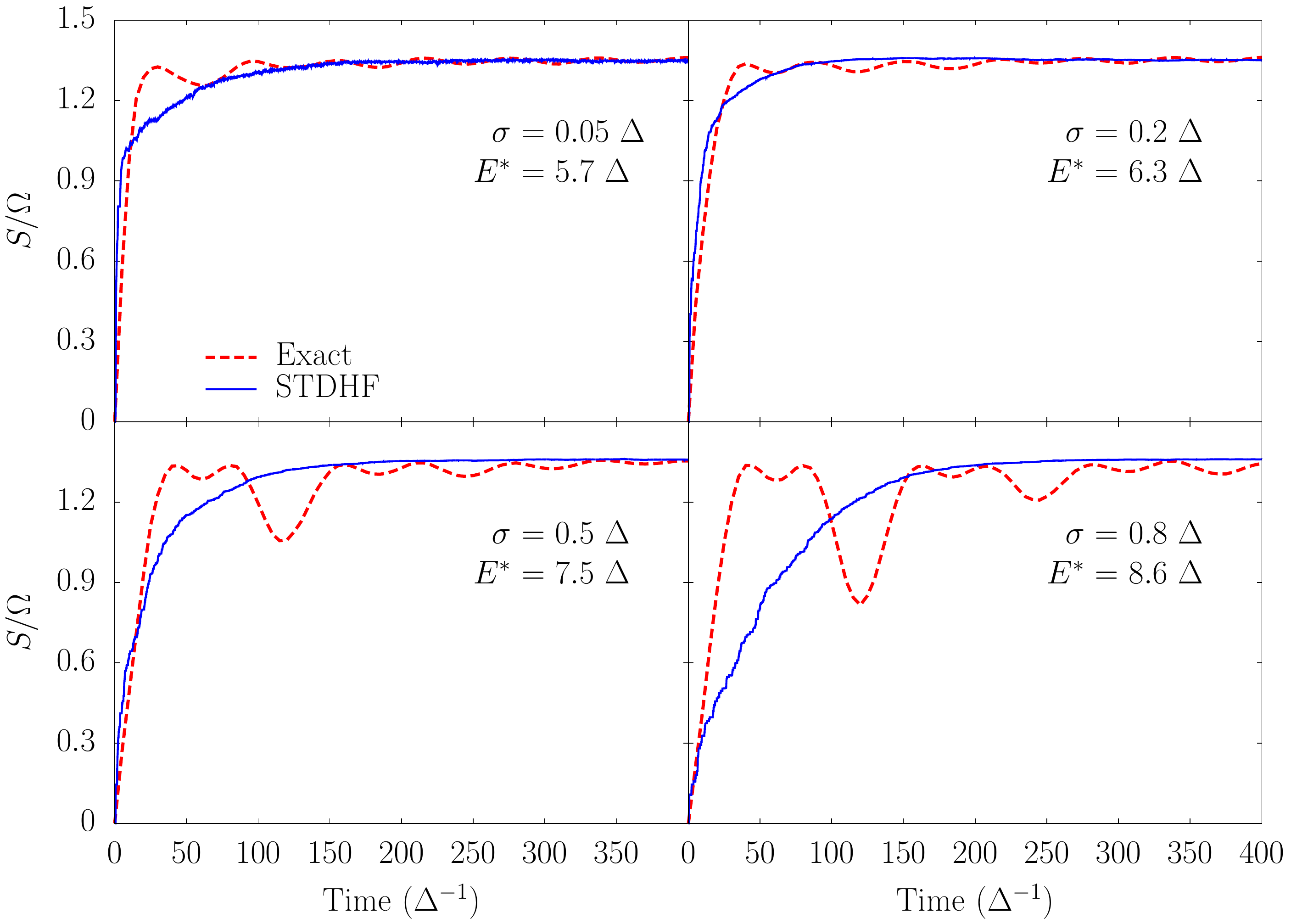}
\caption{\label{fig:entropy_sigma} Entropy per particle $S/\Omega$ as
  a function of time as obtained from exact solution (dashed lines)
  and STDHF (full lines). The four panels show results for four
  different values of band width $\sigma$ as indicated. The other
  parameters are~: $\MD{\lambda}=0.8$, $\gamma=0.3$, $\Omega=10$ and
  $v_0=0.05~\Delta$. The corresponding excitation energy $E^*$ depends on
  $\sigma$ and is also indicated in the plots.}
\end{figure}

Next, we explore the effect of band width $\sigma$ by varying it in
four steps~: $\sigma=0.05$, 0.2, 0.5 and 0.8~$\Delta$. The other
model parameters are kept fixed at $\Omega=10$, $\lambda=0.8$ and
$\gamma=0.3$.
Figure \ref{fig:entropy_sigma} compares the time evolution of the entropy
between exact solution and STDHF. 
Decreasing $\sigma$ reduces the
oscillations of the exact entropy and yields generally faster relaxation to
equilibrium (= maximum entropy). This is plausible because smaller
$\sigma$ produce larger density of states which, in turn, enhances the
chances for jumps and thus delivers more dissipation. A larger $\sigma$ spreads
the spectrum instead and dramatically reduces the phase space accessible
to jumps. Only few states remain in communication and with it, only few
frequencies compete which, in turn, leave longer reverberation before
reaching equilibrium. 

STDHF, as expected, is unable to follow the
oscillations. But it nicely reproduces the trend to increasing
relaxation times with increasing $\sigma$. Indeed, for the same reasons as before,
increasing $\sigma$ decreases the probability of $2p2h$ transitions and therefore, provides a slower 
relaxation time of the STDHF entropy as well. The average predictions are
thus still reliable.

\subsection{Varying the number of particles}
\label{sec:varyN}

Variation of $\sigma$ as done in the previous section changes the density
of states together with energy span for the jumps.  We complement that
by varying the particle number $\Omega$.
Here, we keep the ratio $\sigma/\Omega$ constant (at the value of 0.02).
In such a way, the density of states is kept constant.
Increasing $\Omega$ at constant $\sigma$ would amount to increase
the density of states (as one does when increasing $\sigma$ at constant $\Omega$). 
Since the corresponding results (not shown) are very similar to those presented
below, we here focus on  the impact of increasing the number of particles at constant
density of states.

\begin{figure}[htbp]
\includegraphics[width=\linewidth]{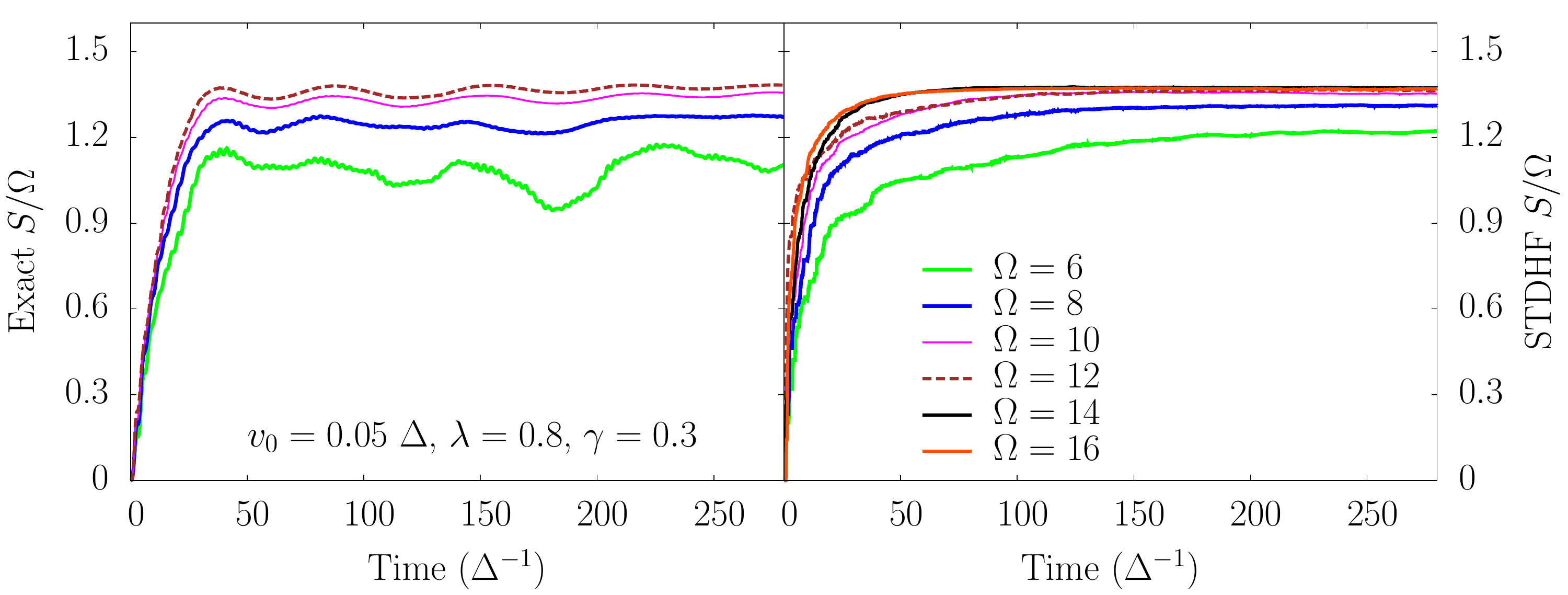}
\caption{\label{fig:entropy_cstdens} Time evolution of the entropy
  per particle $S/\Omega$ for exact solution (left) and STDHF (right)
  and for different numbers of particles $\Omega$ as indicated, while keeping the 
ratio $\sigma/\Omega=0.02$.
  The other model parameters were~: 
  $v_0=0.05~\Delta$, $\lambda=0.8$, $\gamma=0.3$.}
\end{figure}

Figure~\ref{fig:entropy_cstdens}
shows the time evolution of the entropy for the
exact solution (left panel) and for STDHF (right panel).  
We start at
$\Omega=6$ because STDHF does not become active for smaller
$\Omega$. At the upper end, we go up to $\Omega=12$ for the exact
calculation and checked two higher values of $\Omega$ for STDHF.  As
we could expect from the previous results, the amplitude of
oscillations of the exact entropy indeed shrinks with
increasing $\Omega$ since the number of states also increases. We also see
a faster relaxation for increasing $\Omega$, again related to
increasing number of states. This trend to decreasing relaxation time
with increasing $\Omega$ is also reproduced by STDHF.  
We therefore expect that the larger the $\Omega$, the better the agreement between
the exact and the STDHF dynamics. We also observe a convergence of the STDHF results at $\Omega=14$ since 
the time evolution of the entropy changes very little when going from $\Omega=14$ to $\Omega=16$.

There is a
further interesting feature. The entropy indicates two phases of
relaxation. It starts with a fast relaxation at early times and bends
over to a different rate at around 20--30 $\Delta^{-1}$.  This property
is probably also present in the exact
solution although masked by the oscillations.


\section{Conclusion and perspectives}
\label{sec:conclusion}

We have investigated the Stochastic Time-Dependent Hartree-Fock 
(STDHF) approach in a schematic model which allows a comparison with the
exact solution.  The model consists of $N$ fermions in two bands of
single-particle (s.p.) levels, the lower one fully occupied and the
upper one empty. This is augmented by a two-body interaction, which is
motivated first from a typical pairing interaction
and modified by
gradually mixing more and more couplings between different
s.p. levels. The model has much in common with quasi-spin models
consisting out of $N$ coupled spin 1/2 systems, which are widely used
in many areas of physics. The actual tests here are run in the regime
of small interaction strength to maintain a clear hierarchy of
s.p. motion and two-body collisions. For the ground state, we thus see
little difference between Hartree-Fock and exact solution
because  static correlations are weak.
The other model parameters are the number of
particles, the band gap, the band width, and the mixing of transitions in the
two-body interaction. The band gap is taken as the unit of energy and all
other energies count relative to it. The number of particles divided by the
band gap yields the density of states which is a crucial parameter
determining the amount of dissipation by dynamical correlations. 
The dynamical evolution is initialized through an instantaneous boost by a
"dipole'' operator, which in the schematic model is a one-body
operator collecting coherently all vertical transitions between the
two bands. The strength of boost determines the excitation energy
which is one further crucial parameter for the amount of dissipation.

Although the model is built to be very simple, the exact solution
limits the affordable system size. Actually we could perform tests up
to 12 particles. On the other hand, a dissipative model as STDHF
relies on a high density of states to achieve a good mixing of
frequencies and so to justify the Markovian approximation implied in
the concept of instantaneous jumps between two-particle
configurations. The limited particle number means that we are
exploring STDHF in a critical regime and that the situation may be
more forgiving in large systems. 
Even for this limited system size, we
find that STDHF provides a substantial improvement as compared to mere
time-dependent Hartree-Fock (TDHF): STDHF drives convergence to the
same final one-body state as the exact solution and the relaxation
rate is of the correct order. A major difference remains concerning
memory effects. STDHF ignores by construction (Markovian
approximation)  memory effects. The exact solution, however, shows
such memory effects which, e.g., lead to oscillations of the one-body
entropy.  Varying the model parameters allowed to explore the
dependence of memory effects on dynamical regime and system
properties.  Memory effects fade away with increasing excitation
energy, and increasing density of states. 
In situations where memory
effects are small, we find a good agreement of STDHF with the exact
solution. If some oscillations remain in the time evolution of the
entropy, we see at least that STDHF is still providing a reliable
picture of the general trend and of the final state. Thus the result
is very encouraging on the one hand, but also sets a warning flag
which reminds us to check Markovian approximation in each new
situation. The case of non half-filled systems also opens the road for
new aspects, especially at the side of the time evolution of the one-body entropy.
Work in that direction is in progress.

\bigskip

\section*{Acknowledgments}
This work was supported by the CNRS and the Midi-Pyr\'en\'ees region (doctoral allocation number 13050239), and the 
Institut Universitaire de France. It was granted access to the HPC resources of IDRIS under the allocation 2014--095115
made by GENCI (Grand Equipement National de Calcul Intensif), and of CalMiP (Calcul en Midi-Pyr\'en\'ees) 
under the allocation P1238.


\end{document}